\documentclass[aps,prl,twocolumn,superscriptaddress,showpacs]{revtex4-1}
\usepackage{amsmath,amssymb,graphicx,subfigure}

\usepackage[utf8]{inputenc}
\usepackage[T1]{fontenc}
\usepackage{xcolor}

\IfFileExists{newtxtext.sty}
   {\usepackage{newtxtext,newtxmath}}
   {\IfFileExists{stix.sty}
      {\usepackage{stix}}
      {\IfFileExists{mathptmx.sty}
      {\usepackage{mathptmx}}{} } }

\usepackage{textcomp}

\usepackage{bm}

\IfFileExists{siunitx.sty}{\usepackage{booktabs,siunitx}}{}

\pdfoutput=1
\usepackage{color}
\definecolor{LinkColor}{rgb}{0.256,0.439,0.588}
\usepackage{hyperref}
\hypersetup{
   colorlinks=true,
   citecolor=LinkColor,
   linkcolor=LinkColor,
   urlcolor=LinkColor
}

\renewcommand{\vec}[1]{\mathbf{#1}}
\newcommand{\bra}[1]{\langle#1\rvert}
\newcommand{\ket}[1]{\lvert#1\rangle}

\newcommand{\be}{\begin{equation}}
\newcommand{\ee}{\end{equation}}
\newcommand{\bea}{\begin{eqnarray}}
\newcommand{\eea}{\end{eqnarray}}

\usepackage{pifont}

\begin{document}

\title{Pfaffian formalism for higher-order topological insulators }

\author{Heqiu Li}
\affiliation{Department of Physics, University of Michigan, Ann Arbor, MI 48109, USA}
\author{Kai Sun}
\affiliation{Department of Physics, University of Michigan, Ann Arbor, MI 48109, USA}

\date{\today}

\begin{abstract}
We generalize the Pfaffian formalism, which has been playing an important role in the study of time-reversal invariant topological insulators (TIs), to 3D chiral higher-order topological insulators  (HOTIs) protected by the product of four-fold rotational symmetry $C_4$ and the time-reversal symmetry $T$. This Pfaffian description reveals a deep and fundamental link between TIs and HOTIs, and allows important conclusions about TIs to be generalized to HOTIs. As examples, we demonstrate in the Letter how to generalize Fu-Kane's parity criterion for TIs to HOTIs, and also present a general method to efficiently compute the $Z_2$ index of 3D chiral HOTIs without a global gauge.
\end{abstract}

\maketitle

\emph{Introduction.}--- In comparison to the well-studied topological insulators (TIs), which have a gapped $d$-dimensional bulk and topologically-protected gapless states on its $d-1$ dimensional boundaries~\cite{Qi_TI,Qi2008,Hasan_TI,FuKane_Z2,FuKaneMele3D,FuKane_inv,KaneMele2005}, the recently proposed higher-order topological insulators (HOTIs) have a similar gapped bulk, but the gapless states emerge at lower dimensions~\cite{Bernevig_HOTI,Ahn2018,Wieder2018,Ezawa2018,Ezawa2018b,Van2018,Khalaf2018,Kooi2018,Calugaru2019,Benalcazar2017,Benalcazar2017b,Varjas2015,Ezawa2019,
Wang2018,Song2017,Matsugatani2018,Langbehn2017,Yue2019,Hsu2018,Queiroz2018,Xue2019,Schindler2018,Ghorashi2019}, e.g. the 1D hinge of a 3D insulator. In this Letter, we focus on second-order topological insulators characterized by nontrivial magneto-electric polarization $P_3$, e.g., 3D chiral second-order topological insulators (CSOTIs) with gapless chiral hinge states propagating in alternative directions. The physical meaning of this topological invariant can be understood by the theory of electric multipole moments~\cite{Benalcazar2017,Benalcazar2017b}. These second-order TIs have a strong connection to TIs, and in particular, if the time-reversal symmetry $T$ is enforced, $2P_3$ recovers the $Z_2$ index of a TI~\cite{Wang2010}. If the time-reversal symmetry is broken, $2P_3$ still defines a $Z_2$ topological index, as long as a space inversion, rotoinversion or $C_nT$ symmetry is preserved~\cite{Bernevig_HOTI,Ahn2018,Fang2015,Wieder2018,Ezawa2018,Ezawa2018b,Van2018,Khalaf2018,Kooi2018,Lee2019}, where $C_n$ represents $n$-fold rotation with $n=2,4,6$, and this $Z_2$ index, in the absence of time-reversal symmetry, characterizes a second-order TI. For systems invariant under space-inversion or some rotoinversion, this topological index is fully dictated by high-symmetry momenta~\cite{Topchem,indicator,S4index,Ono2018,Kruthoff2017,Slager2012}. However, in general, the diagnosis of higher-order topology requires more sophisticated techniques like the nested Wilson loops \cite{Bernevig_HOTI,Benalcazar2017,Benalcazar2017b,Yu2011,Franca2018,Adrien2018}.

Although TIs and these second-order TIs are characterized by the same $P_3$, which suggests a strong and deep connection between the two, one important link between TIs and second-order TIs is still missing, i.e., the Pfaffian formula for TIs developed by Fu and Kane~\cite{FuKane_Z2}. This Pfaffian formula laid the foundation for many other important conclusions about TIs. For example, in principle, to compute the $Z_2$ index for a TI, it requires global information about the entire Brillouin zone (BZ). In practice,  this means that a global gauge will be needed, such that wavefunctions are globally smooth and continuous in the entire Brillouin zone. Although the existence of such a gauge is guaranteed, finding it is not always straightforward. Based on the Pfaffian formula, several shortcuts were developed to bypass this complicated procedure of finding a global gauge, such as Fu-Kane's high-symmetry point approach for systems with space-inversion symmetry~\cite{FuKane_Z2}, and numerical techniques by Fukui and Hatsugai~\cite{Fukui2007} and by Soluyanov and Vanderbilt \cite{Vanderbilt2011} which dramatically reduced the computational costs. For second order TIs, however, due to the broken time-reversal symmetry, a Pfaffian formalism is still absent, and thus many knowledge that we accumulated from studying TIs cannot be directly generalized.


In this Letter, we develop a Pfaffian formalism for higher-order topological insulators, more precisely CSOTIs, utilizing a composite operator obtained from $C_4T$ sewing matrix. We found that in strong analogy to TIs, the topological index of CSOTIs can also be determined via a Pfaffian formula. This conclusion not only provides a new pathway for computing topological indices, but also makes it possible to generalize existing Pfaffian-based knowledge about TIs to high-order TIs, such as methods to obtain topological indices without a global gauge. As examples, we will show below that our Pfaffian formula provides a straightforward generalization of the Fu-Kane's parity criterion~\cite{FuKane_inv} to second order TIs if a four-fold rotoinversion symmetry is present, which demonstrates a direct connection between $P_3$ and symmetry indicators~\cite{indicator,S4index}. For general CSOTIs without  rotoinversion symmetry, our Pfaffian formalism indicates that high symmetry points alone do not contain sufficient information to fully dictate the topological index, but the Pfaffian formalism allows us to get the index through examining only a small part of the Brillouin zone without using a global gauge, along a similar line as what has been achieved for TIs~\cite{Fukui2007, Vanderbilt2011}.

\emph{Generalization of the Pfaffian formalism.}--- We consider a CSOTI invariant under $C_4T$ but without $T$ or $C_4$ symmetry, and we set the rotational axis to be aligned with the $z$ direction. The more generic systems will be covered in the discussion. 
The half-integer spin leads to $(C_4T)^4=-1$, instead of $(C_4T)^2=-1$ which has been studied in Ref~\onlinecite{Zhang2015}.
Due to the anti-unitary nature of $C_4T$ and the half-integer spin of fermions, in analogy to Kramers doublets, all bands in our system shall show two-fold degeneracy at $C_4T$-invariant momenta, denoted as $K^4=\{\Gamma,M,Z,A\}$, where $\Gamma=(0,0,0),M=(\pi,\pi,0),Z=(0,0,\pi),A=(\pi,\pi,\pi)$.
Without losing generality, we assume that there is no accidental degeneracy beyond what is required by these Kramers pairs, because accidental degeneracy can always be lifted by perturbations without changing topological indices. Thus, for a system with $2N$ valance bands, a $2N\times 2N$ unitary sewing matrix for the symmetry operator $C_4T$ can be defined
\begin{align}
B_{mn}(\mathbf{k})=\bra{u_m(C_4T\;\mathbf{k})}C_4T\ket{u_n(\mathbf{k})},
\end{align}
where $m,n$ are valence band indices, $C_4T\; \mathbf{k} \equiv(k_y,-k_x,-k_z)$ and $\ket{u_n(\mathbf{k})}$ is the periodic part of the Bloch wavefunction.
In the absence of accidental degeneracy as assumed above, this $B(\mathbf{k})$ matrix is $2\times 2$ block diagonal due to the Kramers pairs, i.e., $B=diag(B_1,B_2,...,B_N)$ with $B_r$s being $2\times2$ unitary matrices for $r=1,2,...,N$. According to Ref.~\cite{Wang2010}, there must exist a smooth gauge in our system such that $B_r(\mathbf{k})$ is globally smooth and $\det[B_r(\mathbf{k})]=1$.
Therefore as a function of momentum, each $B_r(\mathbf{k})$ defines a smooth mapping from the 3D BZ to the linear space formed by all $SU(2)$ matrices. In the language of differential manifold, a 3D BZ is a three-torus $T^3$, while $SU(2)$ is
diffeomorphic to a three-sphere $S^3$, and thus $B_r$ defines a mapping  $T^3\rightarrow S^3$. For such a mapping, there exists an integer topological index, i.e. the degree $\deg[B_r]$, which measures how many times the $T^3$ wraps around the $S^3$:
\begin{align}
\deg[B_r]=-\int
\frac{\mathrm{d}^{3} \mathbf{k} }{24 \pi^{2}}\epsilon^{i j k} \operatorname{Tr}\left[\left(B_r \partial_{i} B_r^{\dagger}\right)\left(B_r \partial_{j} B_r^{\dagger}\right)\left(B_r \partial_{k} B_r^{\dagger}\right)\right] \nonumber
\end{align}
where $\partial_i=\partial/\partial{k_i}$.

The definition of magneto-electric polarization $P_3$ can be found in Refs.~\onlinecite{Wang2010,Bernevig_HOTI,Ahn2018,Qi2008,Qi2009,Fang2012,Essin2009,Nicodemos2018}
and it is known that $P_3$ can be computed via the sewing matrix~\cite{Bernevig_HOTI,Fang2012,Ahn2018}
\begin{align}
2 P_{3}=-\frac{1}{24 \pi^{2}} \int \mathrm{d}^{3} \mathbf{k} \epsilon^{i j k} \operatorname{Tr}\left[\left(B \partial_{i} B^{\dagger}\right)\left(B \partial_{j} B^{\dagger}\right)\left(B \partial_{k} B^{\dagger}\right)\right].  \nonumber
\end{align}
For a block diagonal $B$ matrix, this integral reduces to
\begin{align}
2 P_{3}=\sum_{r=1}^N \deg[B_r]
\end{align}
where $\deg[B_r]$ is the degree of the mapping $B_r: T^3\rightarrow S^3$ discussed above.
It is worthwhile to emphasize that only the module 2 of $2 P_3$ (or $\deg[B_r]$) is gauge invariant and thus has real physical meaning. This conclusion can be easily checked by noticing that a gauge transformation can change the degree by an even integer, i.e. under $\ket{u_n(\mathbf{k})}\rightarrow \ket{u_j(\mathbf{k})}U_{jn}(\mathbf{k})$, $B(\mathbf{k})\rightarrow U^\dagger(C_4T\mathbf{k})B(\mathbf{k})U^*(\mathbf{k})$ and $\deg[B]\rightarrow \deg[B]+2 \deg[U]$. Therefore we will only keep track of the mod 2 of the degree, which will be denoted as $\deg_2[B_r]$ in the rest part of this Letter.

The mod 2 of the degree can be easily calculated through a counting technique, if we realize that the degree counts how many times the original spaces wraps around the target space.  Here, we first demonstrate this technique using a simple example: a mapping between 1-spheres $f:S^1\rightarrow S^1$ shown in Fig.~\ref{degree}. To get $\deg_2[f]$, we take any non-singular point in the target space and count how many points in the original space are mapped to this target point under $f$. If this number is $n$, then $\deg_2[f]=n\ mod\ 2$.

\begin{figure}
\includegraphics[width=2.0 in]{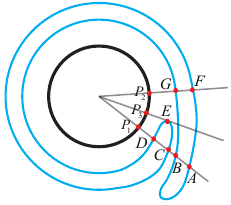}
\caption{ Illustration of a map $f:S^1\rightarrow S^1$ with degree 0. The black circle is the target space and the blue line demonstrates the mapping from the original space to this circle. To calculate $\deg_2[f]$, we can pick a non-singular point like $P_1$ or $P_2$ in the target space and count the number of points that are mapped to it. There are four points mapped to $P_1$ and two points to $P_2$, therefore $\deg_2[f]=4\ mod\ 2=2\ mod\ 2=0$. Note that we cannot choose $P_3$ to calculate the degree because the map at point $E$ is singular.}
\label{degree}
\end{figure}

For $B_r:T^3\rightarrow S^3$, it turns out that a specific gauge can be chosen, which allows the counting technique to be easily adopted.  Because $C_2=-(C_4T)^2$, $C_2$ is also a symmetry of the system, and therefore the $2N\times 2N$ sewing matrix $D_{mn}(\mathbf{k})=\bra{u_m(C_2\mathbf{k})}C_2\ket{u_n(\mathbf{k})}$ is unitary, where $C_2\mathbf{k}\equiv (-k_x,-k_y,k_z)$. The fact that $C_2=-(C_4T)^2$ and $(C_2)^2=-1$ implies~\cite{supp}:
\bea
D(\mathbf{k})=-B(C_4T\mathbf{k})B^*(\mathbf{k})  \;\;\;\textrm{and} \;\;\; D(C_2\mathbf{k})=-D^\dagger(\mathbf{k})
\label{sewing}
\eea
Because $C_2$ does not give rise to nontrivial topology in the presence of the $C_4T$ symmetry, there should be no topological obstruction to smoothly deform the $C_2$ sewing matrix $D(\mathbf{k})$ to a constant matrix independent of momentum $\mathbf{k}$. In the supplemental material~\cite{supp} we explicitly construct a smooth gauge transformation to make $D(\mathbf{k})=diag(i\sigma_z,i\sigma_z,...,i\sigma_z)$ while keeping 
$B_r(\mathbf{k})\in SU(2)$. With this gauge choice, Eq.(\ref{sewing}) implies
\be
B_r(C_4T\mathbf{k})=-i\sigma_z B_r^\mathrm{T}(\mathbf{k})
\label{c4Tfix}
\ee
This condition has remarkable consequences. If $\mathbf{k}$ is a $C_4T$ invariant point, Eq.~\eqref{c4Tfix} implies
\be
B_r(\mathbf{K})=\pm \frac{i}{\sqrt{2}}(\sigma_x+\sigma_y)\equiv A_\pm,\ \mathbf{K}\in K^4,
\label{Apm}
\ee
where $K^4=\{\Gamma,M,Z,A\}$ represents $C_4T$ invariant points as defined early on, i.e., at  $C_4T$ invariant points, $B_r$ can only take one of these two distinct values $A_\pm$. On the other hand, if $\mathbf{k}$ is not a $C_4T$ invariant point and if $B_r(\mathbf{k})=A_\pm$, then Eq.(\ref{c4Tfix}) implies $B_r(C_4T\mathbf{k})=B_r(\mathbf{k})=A_\pm$. Hence if some $\mathbf{k}\notin K^4$ is mapped to $A_+$, there must be one (or three) additional momentum point (related to each other by $C_4T$) which is also mapped to $A_+$, and the same is true for $A_-$. Therefore if we choose $A_+$ (or $A_-$) as the special point to perform the counting described above, as far as $\deg_2[B_r]$ is concerned, only the four $C_4T$ invariant points need to be considered, because any other point contributes even numbers to the counting.  In summary, for each $B_r$, we only need to examine the four $C_4 T$ invariant points ($K^4$). If $n_r$ of these four points are mapped to $A_+$ (and thus $4-n_r$ to $A_-$), then $\deg_2[B_r]=n_r\ mod\ 2$.

\begin{figure}
\includegraphics[width=\columnwidth]{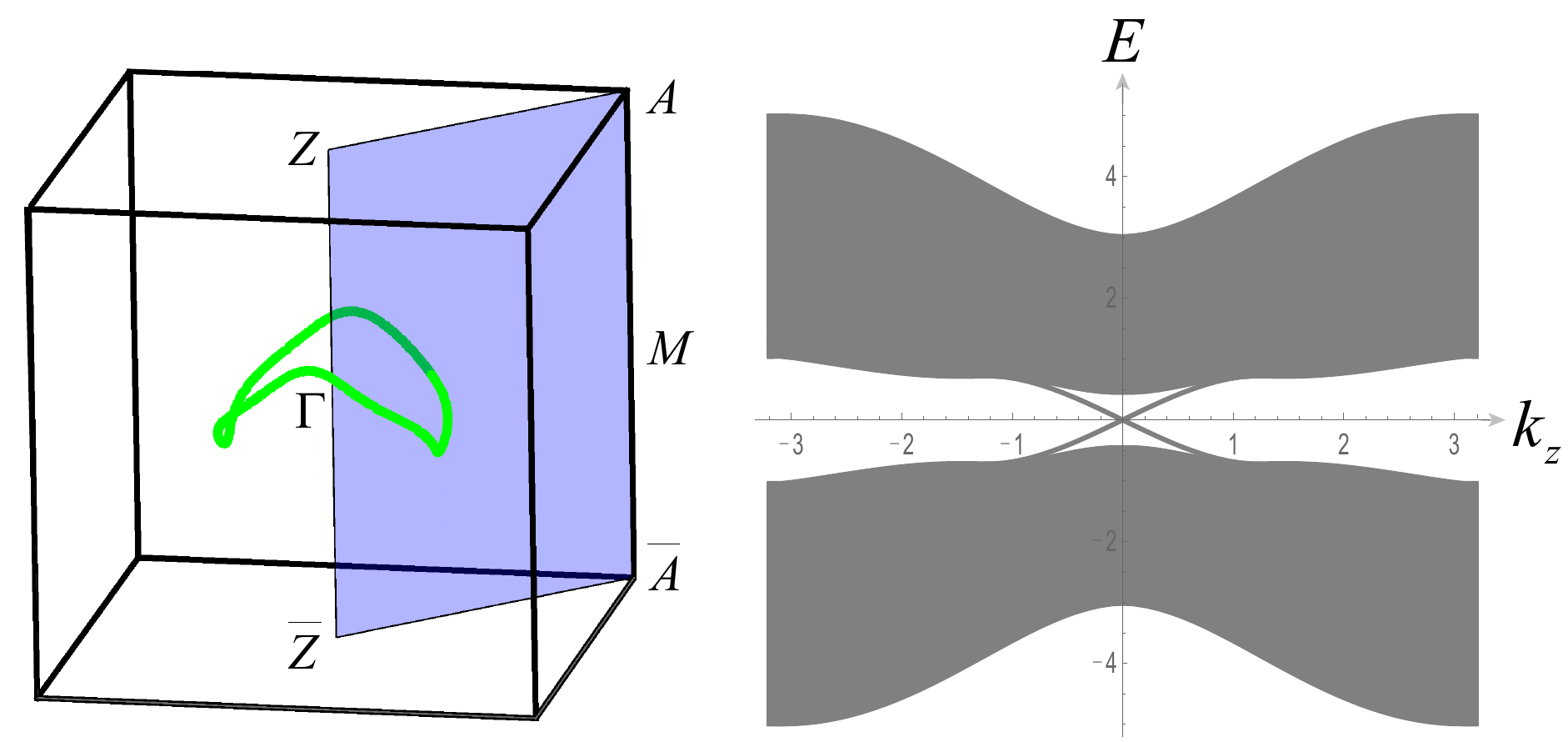}
\caption{ Left: schematic plot of the Brillouin zone. $\tau_{1/2}$ is the colored rectangle and $\partial\tau_{1/2}$ is its boundary. The green line represents the momenta with $\operatorname{Pf}[M(\mathbf{k})]=0$ for the Hamiltonian shown in Eq.~\eqref{H1}. The line of zero Pfaffian penetrates $\tau_{1/2}$, resulting in a phase winding of $2\pi$ in Eq.~\eqref{P3loop}, which implies that $P_3=1/2$. Parameters used here are $p=0.5,q_1=1,q_2=0.2,q_3=0.3$. Right: spectra of $H$ as a function of $k_z$ with open boundary condition along $k_x$ and $k_y$ with the same parameters. The existence of gapless hinge states suggests nontrivial topology.    }
\label{combined}
\end{figure}

Now we relate this $n_r$ to a Pfaffian. Define
\bea
M_{mn}(\mathbf{k})=\bra{u_m(\mathbf{k})} \Theta\ket{u_n(\mathbf{k})},\ \ \Theta=\frac{C_4T+C_4^{-1}T}{\sqrt{2}},
\label{Thetadef}
\eea
where $m,n$ are valence band indices. Under the gauge chosen above, at $\mathbf{K}\in K^4$, $M(\mathbf{K})$ is block diagonal $M(\mathbf{K})=diag(M_1,M_2,...,M_N)$. Using the facts $(C_4)^2=-(C_4)^{-2}$, $T^2=-1$ and $\left(\frac{C_4+C_4^{-1}}{\sqrt{2}}\right)^2=1$, we show in the supplemental material~\cite{supp} that $M(\mathbf{k})$ is antisymmetric $M^\mathrm{T}(\mathbf{k})=-M(\mathbf{k})$ for every $\mathbf{k}$ that is invariant under $C_4$ rotation (straight lines $A\overline{A}$ and $Z\overline{Z}$ in Fig.(\ref{combined})), and at $C_4T$ invariant points $\mathbf{K}\in K^4$, $M(\mathbf{K})$ is unitary and $M(\mathbf{K})=\left(B(\mathbf{K})-B^\mathrm{T}(\mathbf{K}) \right)/\sqrt{2}$. From Eq.~\eqref{Apm}, we know that for $\mathbf{K}\in K^4$, $B_r(\mathbf{K})=A_+$ or $A_-$ and thus respectively  $\operatorname{Pf}[M_r(\mathbf{K})]=+1$ or $-1$, i.e., counting $n_r$ is identical to counting the number of $C_4 T$ invariant momentum points with $\operatorname{Pf}[M_r(\mathbf{K})]=+1$, i.e., $(-1)^{\deg_2[B_r]}=(-1)^{n_r}=\prod_{\mathbf{K}\in K^4}\operatorname{Pf}[M_r(\mathbf{K})] $, and when contributions from all valence bands are combined together, we have
\begin{align}
(-1)^{2P_3}=\prod_{\mathbf{K}\in K^4}\frac{\operatorname{Pf}[M(\mathbf{K})]}{\sqrt{\det[B(\mathbf{K})]}}
\label{P3pfaffian}
\end{align}
This equation is one main conclusion of this Letter. It generalizes the Pfaffian formula of Fu and Kane~\cite{FuKane_Z2} to systems without time-reversal symmetry, via replacing the $T$ operator with a new combination $\Theta=(C_4T+C_4^{-1}T)/\sqrt{2}$.

On the r.h.s. of Eq.~\eqref{P3pfaffian}, we added by hand a denominator $\sqrt{\det B}$. In the gauge we choose above,
this quantity is unity and thus doesn't contribute anything. However, this denominator is important, because it makes the r.h.s. gauge invariant.
Thus, although our conclusion is based on a specific gauge, it remains valid regardless of gauge choices, as long as $B(\mathbf{k})$ remains smooth and a unique sign is chosen for the square root for a continuous branch of $\sqrt{\det[B(\mathbf{k})]}$, which can always be achieved because $B(\mathbf{k})$ is unitary for every $\mathbf k$.
To demonstrate this gauge invariance, here we perform a generic gauge transformation $\ket{u_n(\mathbf{k})}\rightarrow \ket{u_j(\mathbf{k})}U_{jn}(\mathbf{k})$. Because
$\operatorname{Pf}[BAB^{\mathbf{T}}]=\operatorname{Pf}[A]\det[B]$ and $\det[BAB^{\mathbf{T}}]=\det[A]\det[B]^2$, the gauge transformation implies that
$\operatorname{Pf}[M(\mathbf{K})]\rightarrow \operatorname{Pf}[M(\mathbf{K})]\det[U(\mathbf{K})]^*$ and $\sqrt{\det[B(\mathbf{K})]}\rightarrow \sqrt{\det[B(\mathbf{K})]}\det[U(\mathbf{K})]^*$. Hence the effect of the gauge transformation cancels in Eq.~\eqref{P3pfaffian}.

\emph{3D index as a 2D integral.}--- In this part, we will show that Eq.(\ref{P3pfaffian}) can be expressed as a 2D integral, which greatly reduces computational cost  for evaluating $P_3$, similar to what has been achieved in TIs \cite{Fukui2007,Vanderbilt2011}. We define another matrix for the valence bands
\be
\omega_{mn}(\mathbf{k})=\bra{u_m(\mathbf{-k})} \Theta\ket{u_n(\mathbf{k})}
\ee
with $\Theta$ defined in Eq.~\eqref{Thetadef}.
$\omega(\mathbf{k})$ is not unitary for a generic $\mathbf{k}$, but we are mainly interested in $\omega(\mathbf{k})$ for $\mathbf{k}$ inside the straight line formed by $Z\Gamma \overline{Z}$ and $A M \overline{A}$ (Fig.~\ref{combined}). As shown in the supplemental material~\cite{supp}, along these two lines $\omega(\mathbf{k})$ is unitary and
$\det[\omega(\mathbf{k})]=\det[B(\mathbf{k})]$. Because $\omega(\mathbf{K})=M(\mathbf{K})$ for $\mathbf{K}\in K^4$, Eq.(\ref{P3pfaffian}) can thus be rewritten as
\bea
(-1)^{2P_3}&=& \prod_{\mathbf{K}\in K^4}\frac{\operatorname{Pf}[\omega(\mathbf{K})]}{\sqrt{\det[\omega(\mathbf{K})]}}
\label{P3omega}
\eea

As Fu and Kane \cite{FuKane_Z2} have shown for TIs, Eq.(\ref{P3omega}) can also be expressed as an integral
\be
2P_3=\frac{1}{2 \pi}\left[\oint_{\partial \tau_{1 / 2}} A d \ell -\int_{\tau_{1 / 2}} d \tau F\right]\  mod\ 2
\label{halfintegral}
\ee
Here $\tau_{1/2}$ refers to the rectangle $Z\overline{Z}\overline{A}A$ in Fig.~\ref{combined} and $\partial \tau_{1 / 2}$ is its boundary. $A$ and $F$ are the abelian Berry connection and Berry curvature inside $\tau_{1/2}$. If we label each wavefunction $\ket{u_n(\mathbf{k})}$ as $\ket{u_r^{s}(\mathbf{k})}$, where $r=1,...,N$ labels different Kramers pairs and $s=\mathrm{I},\mathrm{II}$ distinguishes the two states in a Kramers pair, there is a gauge fixing condition at the boundary $Z\overline{Z}$ and $A\overline{A}$ for Eq.(\ref{halfintegral}) to be valid:
\bea
\left|u_{r}^{\mathrm{I}}(-\mathbf{k})\right\rangle&=&\Theta\left|u_{r}^{\mathrm{II}}(\mathbf{k})\right\rangle
\label{numericalfix1}
\\
\left|u_{r}^{\mathrm{II}}(-\mathbf{k})\right\rangle&=&-\Theta\left|u_{r}^{\mathrm{I}}(\mathbf{k})\right\rangle
\label{numericalfix2}
\eea
The formula here is slightly different from the one used in Ref.~\onlinecite{FuKane_Z2}, because $T$ is now replaced by $\Theta=(C_4T+C_4^{-1}T)/\sqrt{2}$. But for Eqs.~\eqref{numericalfix1} and~\eqref{numericalfix2}, because it is evaluated only along $Z\overline{Z}$ and $A\overline{A}$, where $\Theta\mathbf{k}=T\mathbf{k}=-\mathbf{k}$,
the difference between $T$ and $\Theta$ vanishes and thus derivations in Ref~\onlinecite{FuKane_Z2} can be generalized to systems studied here by simply replacing $T$ by $\Theta$.

Eq.(\ref{halfintegral}) enables us to develop efficient numerical techniques to calculate $P_3$ without the need for a global gauge, following similar line of thinking as has been achieved for TIs~\cite{Fukui2007,Vanderbilt2011}. The method proceeds as the following. First, let us select a discrete mesh in $\tau_{1/2}$ and define $Q_{\mu,mn}(\mathbf{k})=\bra{u_m(\mathbf{k})} u_n(\mathbf{k+s_\mu})\rangle$, where $\mu=1,2$ and $s_\mu$ is the mesh step size in the two directions in $\tau_{1/2}$. Apply gauge fixing condition Eqs.~\eqref{numericalfix1} and~\eqref{numericalfix2} to the boundary $\partial\tau_{1/2}$. Let $L_\mu(\mathbf{k})=\det[Q_\mu]/|\det[Q_\mu]|$ and $\tilde{A}_\mu(\mathbf{k})=\ln L_\mu(\mathbf{k})$, $\tilde{F}(\mathbf{k})=\ln \left( L_1(\mathbf{k})L_2(\mathbf{k}+s_1)L_1^{-1}(\mathbf{k}+s_2)L_2^{-1}(\mathbf{k}) \right)$ where the imaginary part of all the logarithm are restricted to $(-\pi,\pi]$. Then $2P_3$ can be calculated through
\be
2P_3=\frac{1}{2 i \pi}\left[\sum_{\mathbf{k} \in \partial \tau_{1/2}} \tilde{A}_{\mu}(\mathbf{k})-\sum_{\mathbf{k} \in \tau_{1/2}} \tilde{F}(\mathbf{k})\right] \ mod\ 2,
\label{P3num}
\ee
where the direction $\mu$ should be along the positive direction of $\partial\tau_{1/2}$. This numerical technique does not require a smooth gauge and is thus convenient to implement. This method has been well-known for 2D and 3D TI, and is now generalized to 3D HOTI without time-reversal symmetry.

\emph{$S_4$ symmetry and high-symmetry points.}--- We show in this part that if the system has a fourfold rotoinversion symmetry $S_4$, in addition to $C_4T$, $P_3$ can be directly obtained by evaluating $S_4$ eigenvalues at high symmetry points. This conclusion is a generalization of the Fu-Kane's parity criterion~\cite{FuKane_inv} to HOTIs, with a key observation that $\tilde S=(S_4+S_4^{-1})/\sqrt{2}$ and $\Theta=(C_4T+C_4^{-1}T)/\sqrt{2}$ can play the role of space inversion $I$ and time reversal $T$ respectively. This correspondence can be seen from the fact that $(\tilde S)^2=1$ and $\tilde S \Theta=IT=S_4C_4T$, which is a consequence of $(S_4)^4=-1$ and $S_4=IC_4^{-1}$. Then the derivations shown in Ref.~\onlinecite{FuKane_inv} remain valid as long as we replace $I$ by $\tilde S$ and $T$ by $\Theta$, leading to an expression for $P_3$~\cite{supp}:
\be
(-1)^{2P_3}=\prod_{\mathbf{K}\in K^4}\prod_{r=1}^N \eta_r^{\mathrm{I}}(\mathbf{K})
\label{P3S4}
\ee
Here $r$ runs over all occupied Kramers pairs, and $K_4$ is the set of $S_4$ invariant points and $\eta_r^\mathrm{I}=\pm 1$ is eigenvalue of $\tilde S=(S_4+S_4^{-1})/\sqrt{2}$. Eq.(\ref{P3S4}) is a generalization of the Fu-Kane parity criterion~\cite{FuKane_inv} to systems with $S_4$ but no inversion symmetry. It is also consistent with results obtained using symmetry indicators~\cite{indicator,S4index}. 

\emph{Zeros of the Pfaffian.}--- Eq.(\ref{P3pfaffian}) also allows us to determine $P_3$ though the zero of $\operatorname{Pf}[M(\mathbf{k})]\equiv p_f(\mathbf{k})$. In this section, we no longer assume $S_4$ symmetry. Because $M(\mathbf{k})$ is antisymmetric at every momentum, its Pfaffian is a well-defined function over the whole BZ. Under a smooth gauge with $\det[B]=1$, $p_f(\mathbf{K})=\pm 1$ at $\mathbf{K}\in K^4$. Hence Eq.(\ref{P3pfaffian}) can be interpreted as the sum of phase change of $p_f(\mathbf{k})$ from $Z$ to $\Gamma$ and from $M$ to $A$ as shown in Fig.~\ref{combined}, i.e., $2P_3=(\pi i)^{-1} \int_{L} d\mathbf{k} \cdot \mathbf{\nabla} \ln p_f(\mathbf{k})  $ where $L$ is the combination of two straight pathes $(Z\rightarrow\Gamma)+(M\rightarrow A)$. As proved in the supplemental material~\cite{supp}, $p_f(\mathbf{k})=p_f(C_4T\mathbf{k})^*\det[B(\mathbf{k})]$, and thus when $\det[B]=1$ and $\mathbf{k}\in L$, $p_f(\mathbf{k})=p_f(-\mathbf{k})^*$. Therefore the phase change of $p_f(\mathbf{k})$ from $Z$ to $\Gamma$ is the same as that from $\Gamma$ to $\overline Z$. With this fact we can extend the integration path $L$ to be $\partial\tau_{1/2}$ and divide by $2$ to get $2P_3$, which gives
\be
2P_3=\frac{1}{2\pi i}\oint_{\partial\tau_{1/2}} d\mathbf{k} \cdot \mathbf{\nabla} \ln \operatorname{Pf}[M(\mathbf{k})]
\label{P3loop}
\ee
This is a generalization of the result by Kane and Mele \cite{KaneMele2005}, via replacing $T$ by $\Theta$. The r.h.s. of the equation measures the phase winding of $\operatorname{Pf}[M(\mathbf{k})]$ around the boundary of the 2D area $\tau_{1/2}$. Because a nontrivial phase winding around the 1D boundary implies nodal points in the 2D bulk
 with $\operatorname{Pf}[M(\mathbf{k})]=0$, this equation implies that $2 P_3$ can be obtained by counting the number of nodal points with $\operatorname{Pf}[M(\mathbf{k})]=0$ in $\tau_{1/2}$. More details will be demonstrated
below using a tight-binding model. Interestingly, here we have shown that for a 3D HOTI, its topological index $P_3$ can be calculated by looking at the zeros of $\operatorname{Pf}[M(\mathbf{k})]$ in a single 2D plane ($\tau_{1/2}$). This is in direct contrast to first order 3D TIs, where one needs to investigate two time reversal invariant 2D planes to determine the $Z_2$ index~\cite{FuKaneMele3D}. Eq.(\ref{P3loop}) also implies that if $\operatorname{Pf}[M(\mathbf{k})]$ is nonzero over $\tau_{1/2}$, then $P_3$ will automatically be trivial.

\emph{Tight binding model.}--- Here we use tight-binding models to demonstrate and to verify our conclusions. Consider a four-band model with a Hamiltonian
\bea
H(\mathbf{k})&=&(\cos k_x+\cos k_y+\cos k_z-2)\tau_z\sigma_0+q_1 \sum_{i=x,y,z} \sin k_i\tau_x\sigma_i \nonumber\\
&&+ q_2 \sum_{j=x,y} \sin k_j \sin k_z\tau_y\sigma_j+q_3 \tau_x\sigma_0 \nonumber\\
&&+ p (\cos k_x-\cos k_y)\tau_y\sigma_0
\label{H1}
\eea
Here $C_4=\tau_0 e^{-i\frac{\pi}{4}\sigma_z}$, $T=-i\tau_0\sigma_yK$, $S_4=\tau_z e^{i\frac{\pi}{4}\sigma_z}$. The Hamiltonian satisfies $C_4T H(\mathbf{k}) (C_4T)^{-1}=H(C_4T\mathbf{k})$.   The $p$ term breaks $C_4$ and $T$ symmetry but preserves $S_4$ and $C_4T$. If $p$ vanishes then the system becomes a 3D TI. The $q_2$ and $q_3$ terms break $S_4$ symmetry. When $q_2=q_3=0$, $S_4$ symmetry is recovered and the model reduces to the one shown in Ref.~\onlinecite{Bernevig_HOTI}. In this case the Kramers pair in the valence bands at $\Gamma$ has $\tilde S$ eigenvalue $-1$ and all other $S_4$ invariant points have $\tilde S$ eigenvalue $+1$. Thus by Eq.~\eqref{P3S4}, we have $P_3=1/2$ and the system is a CSOTI. When small $q_2$ and $q_3$ are turned on, the band gap does not close and the system should still remain a CSOTI. We calculate the zero of $\operatorname{Pf}[M(\mathbf{k})]$ as shown in Fig.~\ref{combined}. The zeros form a loop penetrating $\tau_{1/2}$, giving rise to a phase winding of $2\pi$ in $\operatorname{Pf}[M(\mathbf k)]$. Therefore from Eq.~\eqref{P3loop}, $P_3=1/2$. We also apply Eq.~\eqref{P3num} and get $P_3=1/2$ as well. To verify our prediction we diagonalize the system with open boundary condition along $k_x$ and $k_y$, and the spectra as a function of $k_z$ is shown in Fig.~\ref{combined}. Gapless hinge states are found, which confirms that the system is a CSOTI with $P_3=1/2$.


\emph{Conclusions and discussions.}---In this Letter, we generalize the Pfaffian topological invariant to higher-order topological insulators, utilizing a composite operator composed of linear superposition of symmetry operators. In addition to the $C_4T$-invariant systems discussed above,  this construction applies generically to systems with symmetry-enforced Kramers-like degeneracy (see S-6 in the SI~\cite{supp} for more details). In addition, this Pfaffian formula is directly related with the dipole pumping and the nontrivial Wannier-band Chern numbers~\cite{Benalcazar2017,Benalcazar2017b}, as shown in S-7 in the SI~\cite{supp}).

\emph{Acknowledgement.}--- H.L. and K.S. acknowledge the support of the National Science Foundation Grant No. EFRI-1741618.


%

\appendix

\begin{widetext}

\section{Supplemental Material for "Pfaffian formalism for higher-order topological insulators"}

\section{Properties of matrices}

In the main text and supplemental material we defined many matrices using symmetry operators. Here we prove some of their properties that are important for the proof in the main text. The definitions are:
\bea
B_{mn}(\mathbf{k})&=&\bra{u_m(C_4T\mathbf{k})}C_4T\ket{u_n(\mathbf{k})} \\
D_{mn}(\mathbf{k})&=&\bra{u_m(C_2\mathbf{k})}C_2\ket{u_n(\mathbf{k})}  \\
M_{mn}(\mathbf{k})&=&\bra{u_m(\mathbf{k})} \Theta\ket{u_n(\mathbf{k})},\ \ \Theta=\frac{C_4T+C_4^{-1}T}{\sqrt{2}} \\
\omega_{mn}(\mathbf{k})&=&\bra{u_m(\mathbf{-k})} \Theta\ket{u_n(\mathbf{k})} \\
v_{mn}(\mathbf{k})&=&\bra{u_m(\mathbf{k})}S_4C_4T\ket{u_n(\mathbf{k})}
\eea
All of these matrices are $2N\times 2N$ where $2N$ is the number of occupied bands.

Proof of unitarity: in general if a gapped system has a symmetry $P$ that is either unitary or antiunitary so that $PH(\mathbf k)P^{-1}=H(P\mathbf k)$, then $H(P\mathbf k)P\ket{u_n(\mathbf k)}=PH(\mathbf k)\ket{u_n(\mathbf k)}=E_n(\mathbf k)P\ket{u_n(\mathbf k)}$ so that $P\ket{u_n(\mathbf k)}$ is a superposition of eigenstate of $H(P\mathbf k)$ with the same energy. If $\ket{u_n(\mathbf k)}$ is in occupied bands, so should $P\ket{u_n(\mathbf k)}$, then we have $P\ket{u_n(\mathbf k)}=\sum_{m\in occ}\ket{u_m(P\mathbf k)}\tilde{P}_{mn}(\mathbf k)$ where $\tilde{P}_{mn}(\mathbf k)$ is the coefficient for the superposition. Since $\tilde{P}_{mn}(\mathbf k)$ is now the transformation matrix between two sets of normalized and orthogonal basis $\{ P\ket{u_n(\mathbf k)} \}$ and $\{ \ket{u_n(P\mathbf k)} \}$ that span the space of occupied bands, $\tilde{P}_{mn}(\mathbf k)=\bra{u_m(P\mathbf k)}P\ket{u_n(\mathbf k)}$ is guaranteed to be unitary.

This fact shows immediately that matrices $B$, $D$ and $v$ above are unitary for every $\mathbf k$ as long as the system has the corresponding symmetry. $M(\mathbf k)$ is unitary only at $C_4T$ invariant points since $C_4T\mathbf k=C_4^{-1}T\mathbf k=\mathbf k$ there. $\omega(\mathbf k)$ is unitary at straight lines $Z\overline{Z}$ and $A\overline{A}$ since $C_4T\mathbf k=C_4^{-1}T\mathbf k=-\mathbf k$ there.

Proof of Antisymmetry: we will show that $M$ is antisymmetric for every $\mathbf k$ that is invariant under $C_4$, such that $C_4T\mathbf k=C_4^{-1}T\mathbf k=-\mathbf k$. First we show that $\Theta^2=T^2=-1$. Since $(C_2)^2=-1$, then $C_2+C_2^{-1}=0$ and $\Theta^2=-(C_4+C_4^{-1})^2/2=-(C_2+C_2^{-1}+2)/2=-1$.
\bea
M_{mn}(\mathbf{k})&=&\bra{u_m(\mathbf{k})} \Theta\ket{u_n(\mathbf{k})} \nonumber\\
&=&\bra{\Theta^2 u_n(\mathbf{k})} \Theta\ket{u_m(\mathbf{k})}\nonumber\\
&=&-\bra{u_n(\mathbf{k})} \Theta\ket{u_m(\mathbf{k})}\nonumber\\
&=&-M_{nm}(\mathbf{k})
\eea
Similarly $v_{mn}(\mathbf{k})=-v_{nm}(\mathbf{k})$ so that $v$ is also antisymmetric. The same technique when applied to $\omega$ and $B$ yields
\bea
\omega_{mn}(\mathbf{k})&=&-\omega_{nm}(\mathbf{-k})  \nonumber\\
 B_{mn}(\mathbf{k})&=&-\bra{u_n(\mathbf{k})}C_4^{-1}T\ket{u_m(C_4T\mathbf{k})}
\label{Btranspose}
\eea

Constrains from operator identities: consider the complete relation $\textbf{1}=\sum_{i\in occ}\ket{u_i(\mathbf{k})}\bra{u_i(\mathbf{k})}+\sum_{i\in uocc}\ket{u_i(\mathbf{k})}\bra{u_i(\mathbf{k})}$ where $occ$ and $uocc$ refer to occupied and unoccupied bands. Since $(C_2)^2=-1$, let $\ket{u_m}$ and $\ket{u_n}$ be in occupied bands, replace the $\mathbf{k}$ in the complete relation by $C_2\mathbf{k}$ and insert it to the $\textbf{1}$ in identity $\bra{u_m(\mathbf{k})} C_2 \textbf{1} C_2 \ket{u_n(\mathbf{k})}=-\delta_{mn}$. The unoccupied part in the complete relation do not contribute since $C_2 \ket{u_n(\mathbf{k})}$ belongs to the space spanned by occupied bands and have zero overlap with unoccupied bands. Therefore we get $\sum_{i\in occ}D_{mi}(C_2\mathbf{k})D_{in}(\mathbf{k})=-\delta_{mn}$, which means
\be
 D(C_2\mathbf{k})=-D^\dagger(\mathbf{k})
 \label{Dcyclic}
\ee
We can also consider identity $C_2=-(C_4T)^2$. Replace the $\mathbf{k}$ by $C_4T\mathbf{k}$ in complete relation and insert it to $\bra{u_m(C_2\mathbf{k})} C_2 \ket{u_n(\mathbf{k})}=-\bra{u_m(C_2\mathbf{k})} C_4T \textbf{1} C_4T \ket{u_n(\mathbf{k})}$. Note that $T$ is antiunitary therefore a complex conjugation is needed, and we get
\be
D(\mathbf{k})=-B(C_4T\mathbf{k})B^*(\mathbf{k})
\label{DBrelation}
\ee
Similarly for operator identity $S_4C_4T=-C_4^{-1}TS_4C_4TC_4T$, replace the $\mathbf{k}$ by $C_4T\mathbf{k}$ in complete relation and insert it to $\bra{u_m(\mathbf{k})} S_4C_4T \ket{u_n(\mathbf{k})}=-\bra{u_m(\mathbf{k})} C_4^{-1}T \textbf{1} S_4C_4T \textbf{1}  C_4T \ket{u_n(\mathbf{k})}$, we get $v_{mn}(\mathbf{k})=-\sum_{i,j\in occ}\bra{u_m(\mathbf{k})} C_4^{-1}T\ket{u_i(C_4T\mathbf{k})} v_{ij}^*(C_4T\mathbf{k})B_{jn}(\mathbf{k})$. Using Eq.(\ref{Btranspose}) we have
\bea
v(\mathbf{k})&=&B^\mathrm{T}(\mathbf{k})v^*(C_4T\mathbf{k})B(\mathbf{k})  \\
\operatorname{Pf}[v(\mathbf{k})]&=&\operatorname{Pf}[v(C_4T\mathbf{k})]^*\det[B(\mathbf{k})]
\eea
If we replace $S_4C_4T$ by $\Theta$, the above derivations are still true and we will get
\bea
M(\mathbf{k})&=&B^\mathrm{T}(\mathbf{k})M^*(C_4T\mathbf{k})B(\mathbf{k})  \\
\operatorname{Pf}[M(\mathbf{k})]&=&\operatorname{Pf}[M(C_4T\mathbf{k})]^*\det[B(\mathbf{k})]
\eea

Determinant equalities: we will show that $\det[B(\mathbf{K})]=\det[M(\mathbf{K})]$ for $\mathbf{K}\in K^4$ and $\det[B(\tilde{k})]=\det[\omega(\tilde{k})]$ for $\tilde{k}$ at straight lines $Z\overline{Z}$ and $A\overline{A}$. Since $C_4T\mathbf{K}=C_4^{-1}T\mathbf{K}=\mathbf{K},\ C_4T\tilde{k}=C_4^{-1}T\tilde{k}=-\tilde{k}$, using Eq.(\ref{Btranspose}) we get
\bea
M_{mn}(\mathbf{K})&=&(\bra{u_m(\mathbf{K})}C_4T \ket{u_n(\mathbf{K})}+\bra{u_m(\mathbf{K})}C_4^{-1}T \ket{u_n(\mathbf{K})})/\sqrt{2}=(B(\mathbf{K})-B^{\mathrm{T}}(\mathbf{K}))/\sqrt{2} \nonumber\\
\omega_{mn}(\tilde{k})&=&(\bra{u_m(-\tilde{k})}C_4T \ket{u_n(\tilde{k})}+\bra{u_m(-\tilde{k})}C_4^{-1}T \ket{u_n(\tilde{k})})/\sqrt{2}=(B(\tilde{k})-B^{\mathrm{T}}(-\tilde{k}))/\sqrt{2}
\label{MomegaB}
\eea
When there is no accidental degeneracy, we can find a gauge where $B(\mathbf{k})$ is block diagonal with each block being $B_r(\mathbf{k})\in SU(2)$. Therefore $M(\mathbf{K})$ and $\omega(\tilde{k})$ are also block diagonal. Since $C_2\tilde k=\tilde k$, Eq.(\ref{Dcyclic}) implies each block of $D$ is $D_r(\tilde{k})=i\hat n(\tilde{k})\cdot \vec\sigma$. Then Eq.(\ref{DBrelation}) gives $B_r(-\tilde{k})=-(i\hat n(\tilde{k})\cdot \vec\sigma) B_r^\mathrm{T}(\tilde{k})$. Let $\hat n'(\tilde{k})=(n_x,-n_y,n_z)$, Eq.(\ref{MomegaB}) leads to
\bea
\omega_r(\tilde{k})&=&B_r(\tilde{k})\frac{1+i\hat n'(\tilde{k}) \cdot \vec\sigma}{\sqrt{2}}  \nonumber\\
\det[\omega_r(\tilde{k})]&=&\det[B_r(\tilde{k})]\det \left[\frac{1+i\hat n'(\tilde{k}) \cdot \vec\sigma}{\sqrt{2}}\right]=\det[B_r(\tilde{k})] \nonumber\\
\det[\omega(\tilde{k})]&=&\det[B(\tilde{k})]
\eea
Now we have proved the $\det[\omega(\tilde{k})]=\det[B(\tilde{k})] $ for the gauge in which $\det[B]=1$. But since $\det[B(\tilde{k})]$ and $\det[\omega(\tilde{k})]$ change in the same way under gauge transformation, they should be equal in any gauge. At $C_4T$ invariant due to $\omega(\mathbf{K})=M(\mathbf{K})$, we also have $\det[B(\mathbf{K})]=\det[M(\mathbf{K})]$.

Properties of the degree of map: we consider a general smooth map $G: T^3\rightarrow SU(2)$. The degree of this map is
\be
\deg [G]= -\frac{1}{24 \pi^{2}} \int \mathrm{d}^{3} \mathbf{k} \epsilon^{i j k} \operatorname{Tr}\left[\left(G \partial_{i} G^{\dagger}\right)\left(G \partial_{j} G^{\dagger}\right)\left(G \partial_{k} G^{\dagger}\right)\right]
\label{degreedef}
\ee
One can show that the degree is quantized to integer and thus invariant for small perturbation of the map \cite{Bernevig_HOTI}. The degree is addictive, which means if there is another map $G': T^3\rightarrow SU(2)$, then
\be
\deg[GG']=\deg[G]+\deg[G']
\ee
Therefore $G^{-1}=G^\dagger$ always has the opposite degree to $G$ since their product is constant. Taking a complex conjugation to Eq.(\ref{degreedef}) we get the degree of $G^*$. Since $\deg[G]$ is real, this means $G^*$ has the same degree as $G$. Combining these facts we get
\bea
\deg[G^\dagger]=-\deg[G] ,\ \ \deg[G^*]=\deg[G]  ,\ \ \deg[G^\mathrm{T}]=-\deg[G]
\eea
If we perform a rotation or rotoinversion in the $\mathbf{k}$ space $T^3$ given by $s:T^3\rightarrow T^3$ so that $s\mathbf{k}=\mathbf{k}'$, the degree of the transformed map $G\circ s$ can be obtained by a change of variable in Eq.(\ref{degreedef}). During the change of variable the volume element $\mathrm{d}^{3} \mathbf{k}$ will bring a factor of $|\det\left[ \frac{\partial (k_x,k_y,k_z)}{\partial (k'_x,k'_y,k'_z)} \right]|$ and the partial derivatives will introduce a factor of $\det\left[ \frac{\partial (k'_x,k'_y,k'_z)}{\partial (k_x,k_y,k_z)} \right]$, therefore the overall effect is to introduce a factor of $\mathrm{sign}\left( \det\left[ \frac{\partial (k'_x,k'_y,k'_z)}{\partial (k_x,k_y,k_z)} \right] \right)$
\bea
\deg [G\circ s]&=& -\frac{1}{24 \pi^{2}} \int \mathrm{d}^{3} \mathbf{k} \epsilon^{i j k} \operatorname{Tr}\left[\left(G(s\mathbf{k}) \partial_{i} G(s\mathbf{k})^{\dagger}\right)\left(G(s\mathbf{k}) \partial_{j} G(s\mathbf{k})^{\dagger}\right)\left(G(s\mathbf{k}) \partial_{k} G(s\mathbf{k})^{\dagger}\right)\right]  \nonumber\\
&=&\mathrm{sign}\left( \det\left[ \frac{\partial (k'_x,k'_y,k'_z)}{\partial (k_x,k_y,k_z)} \right] \right) \deg [G]
\eea
In particular if the transformation $s$ is $C_4T$ so that $\mathbf{k}'=(k_y,-k_x,-k_z)$, then $\det\left[\frac{\partial (k'_x,k'_y,k'_z)}{\partial (k_x,k_y,k_z)}\right] =-1$. Therefore
\be
\deg[G(C_4T\mathbf{k})]=-\deg[G(\mathbf{k})]
\ee

\section{Explicit transformation to make $C_2$ sewing matrix constant}

In the gauge in which the $C_4T$ sewing matrix $B(\mathbf{k})$ is smooth and $\det[B]=1$, the $C_2$ sewing matrix $D(\mathbf{k})$ is also smooth and $\det[D]=1$. We want to find a smooth gauge transformation $\ket{u_n(\mathbf{k})}\rightarrow \ket{u_j(\mathbf{k})}\tilde{U}_{jn}(\mathbf{k})$ where $\tilde{U}\in SU(2N)$ to make $C_2$ sewing matrix $D(\mathbf{k})=diag(i\sigma_z,i\sigma_z,...,i\sigma_z)$ for all momenta in the BZ. Since we assume no accidental degeneracy other than Kramers degeneracy, $D$ is block diagonal with each block being $SU(2)$. Therefore the problem reduce from $SU(2N)$ to $SU(2)$. Denote $D_r(\mathbf{k})$ as one of the $SU(2)$ block, under the gauge transformation $D_r(\mathbf{k})\rightarrow U^\dagger(C_2\mathbf{k}) D_r(\mathbf{k}) U(\mathbf{k})$. Since $D_r\in SU(2)$, we can parameterize it as
\be
D_r(\mathbf{k})=\exp [i(\pi/2+\delta(\mathbf{k})) \hat n(\mathbf{k}) \cdot \vec{\sigma} ],
\ee
where $\delta(\mathbf{k})\in [-\frac{\pi}{2},\frac{\pi}{2}]$ and $\hat n=(n_x,n_y,n_z)$ is a unit vector that can be interpreted as the axis of the $SU(2)$ rotation, and $2(\pi/2+\delta)\in [0,2\pi]$ is the angle of rotation. The fact that $D_r(C_2\mathbf{k})=-D_r^\dagger(\mathbf{k})$ implies
\bea
\delta(\mathbf{k})=-\delta(C_2\mathbf{k}),\ \ \hat n(\mathbf{k})=\hat n(C_2\mathbf{k})
\eea
Because $D_r(\mathbf{k})=-B_r(C_4T\mathbf{k})B_r^*(\mathbf{k})$ and $B_r(C_4T\mathbf{k})$ has the opposite degree to $B_r^*(\mathbf{k})$, the degree of $D_r(\mathbf{k})$ is zero and the winding is trivial. We will first find a transformation to make $\delta(\mathbf{k})=0$. This can be achieved by choosing $R(\mathbf{k})=\exp[-i\frac{1}{2}\delta(\mathbf{k})\hat n(\mathbf{k})\cdot \vec \sigma]$. Then $R(C_2\mathbf{k})=\exp[i\frac{1}{2}\delta(\mathbf{k})\hat n(\mathbf{k})\cdot \vec \sigma]$ and
\bea
 R^\dagger(C_2\mathbf{k})D_r(\mathbf{k})R(\mathbf{k})&=&\exp[-i\frac{1}{2}\delta(\mathbf{k})\hat n(\mathbf{k})\cdot \vec \sigma] \exp [i(\pi/2+\delta(\mathbf{k})) \hat n(\mathbf{k}) \cdot \vec{\sigma} ] \exp[-i\frac{1}{2}\delta(\mathbf{k})\hat n(\mathbf{k})\cdot \vec \sigma]  \nonumber\\
 &=& \exp [i(\pi/2) \hat n(\mathbf{k}) \cdot \vec{\sigma} ] =i\hat n(\mathbf{k}) \cdot \vec{\sigma}
\eea
Now the target space is determined by unit vector $\hat n$ and is isomorphic to a subspace of $S^2$. Since the winding is trivial, there exists a similarity transformation $w(\mathbf{k})\in SU(2)$ to bring each axis $\hat n(\mathbf{k})$ to $+\hat z$ direction:
\be
w^\dagger(\mathbf{k}) ( i\hat n(\mathbf{k}) \cdot \vec{\sigma} ) w(\mathbf{k})=i \sigma_z
\label{axisrotate}
\ee

\begin{figure}
\includegraphics[width=1.8in]{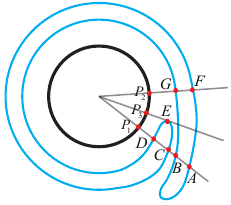}
\caption{ Illustration of a map $f:S^1\rightarrow S^1$ with degree 0. The black circle is the target space and the blue line demonstrates the mapping from the original space to this circle. }
\label{degree}
\end{figure}

See Fig.(\ref{degree}) for an analogy of a map $f:S^1\rightarrow S^1$. The map $f$ has degree $0$, therefore the target of each $\mathbf{k}$ can be continuous deformed to the north pole by either a clockwise deformation (for points C, D, E) or a counter clockwise deformation (for points A, B, F, G). Now we show that $w(\mathbf{k})=w(C_2\mathbf{k})$. Suppose momentum $\mathbf{k_0}$ corresponds to point $A$ in Fig.(\ref{degree}). The fact that $\hat n(\mathbf{k})=\hat n(C_2\mathbf{k})$ means $C_2\mathbf{k_0}$ must be one of $B, C, D$ points so that they have the same $\hat n$. If $C_2\mathbf{k_0}$ corresponds to $C$ or $D$, then if we let $\mathbf{k}$ vary on a continuous path of momentum points from $A$ to $F$, the point $C_2\mathbf{k}$ must have a discontinuous jump at $E$, which is impossible. Therefore $C_2\mathbf{k_0}$ has to be point $B$, which is in the same branch as $A$. Therefore if transformation $w(\mathbf{k_0})$ can bring $A$ to the north pole, so can it bring $B$. Hence we have $w(\mathbf{k})=w(C_2\mathbf{k})$.

Now the transformation to bring $D_r(\mathbf{k})$ to $i\sigma_z$ is just $U(\mathbf{k})=R(\mathbf{k})w(\mathbf{k})$, so that
\bea
D_r(\mathbf{k})&\rightarrow& U^\dagger(C_2\mathbf{k}) D_r(\mathbf{k}) U(\mathbf{k}) \nonumber\\
&=&w^\dagger(C_2\mathbf{k})  R^\dagger(C_2\mathbf{k})D_r(\mathbf{k})R(\mathbf{k}) w(\mathbf{k}) \nonumber\\
&=& w^\dagger(\mathbf{k}) ( i\hat n(\mathbf{k}) \cdot \vec{\sigma} ) w(\mathbf{k})=i \sigma_z
\eea

\section{Issues about singular points in degree counting  }

In the main text when we use the degree counting technique on $C_4T$ invariant points $\mathbf{K}\in K^4$ to calculate the degree of $B_r(\mathbf{k})$, there is a requirement that the map at $\mathbf{K}$ is not singular. If we parameterize the map $B_r(\mathbf{k})=\exp[i\theta \hat n\cdot \vec \sigma]$ with constrain $n_x^2+n_y^2+n_z^2=1$, then the non-singular condition is that the determinant of Jacobian matrix at $\mathbf{K}$ is nonzero: $\det \left[\frac{\partial(\theta,n_\alpha,n_\beta)}{\partial(k_x,k_y,k_z)}\right]\ne 0$ for some $\alpha,\beta\in\{ x,y,z \}$.

In general we can always do a small perturbation to the map to avoid the singular points, unless the singularity is enforced by symmetry. One such example is the sewing matrix $D_r(\mathbf{k})\in SU(2)$ for $C_2$. Symmetry requires $D_r(C_2\mathbf{k})=-D_r^\dagger(\mathbf{k})$ hence $\hat n(C_2\mathbf{k})=\hat n(\mathbf{k})$. Thus near $C_2$ invariant points $\hat n(k_x,k_y,k_z)=\hat n(-k_x,-k_y,k_z)$ for small $k_x,k_y$. Therefore $\frac{\partial\hat n}{\partial k_x}=\frac{\partial\hat n}{\partial k_y}=0$, leading to a zero determinant of Jacobian. This means the map $D_r(\mathbf{k})$ at every $C_2$ invariant point must be singular and the degree counting technique does work for it.

Fortunately there is no such symmetry restriction for $C_4T$ sewing matrix $B_r(\mathbf{k})$. Consider momentum points near $\Gamma=(0,0,0)$. In the gauge where $D(\mathbf{k})=diag(i\sigma_z,...,i\sigma_z)$, symmetry requires
\bea
B_r(C_4T\mathbf{k})&=&B_r(k_y,-k_x,-k_z)=-i\sigma_z B^\mathrm{T}(\mathbf{k}) \nonumber\\
B_r(\Gamma)&=&\pm \frac{i}{\sqrt{2}}(\sigma_x+\sigma_y)
\label{C4Tnonsingular}
\eea
We can give an example of $B_r$ that is non-singular near $\Gamma$ and respect the symmetry requirement. Keep only first order terms in $k$, this example is given by
\be
B_r(\mathbf{k})=k_x+i(\frac{1}{\sqrt{2}}+k_z)\sigma_x+i(\frac{1}{\sqrt{2}}-k_z)\sigma_y+ik_y\sigma_z +O(k^2)
\ee
It can be checked that this example is consistent with Eq.(\ref{C4Tnonsingular}), and the determinant of Jacobian $\det \left[\frac{\partial(\theta,n_x,n_z)}{\partial(k_x,k_y,k_z)}\right]\ne 0$. Therefore there is no symmetry restriction for $B_r$ to be singular at $C_4T$ invariant points and we are legitimate to apply the degree counting technique in the main text.

\section{Calculation of $P_3$ in systems with $S_4$ symmetry  }

In this section we explicitly show that $P_3$ can be obtained from $S_4$ eigenvalues when the system has $S_4$ symmetry in addition to $C_4T$ symmetry. Define sewing matrix
\be
v_{mn}(\mathbf{k})=\bra{u_m(\mathbf{k})}S_4C_4T\ket{u_n(\mathbf{k})}
\ee
As shown above, $v(\mathbf{k})$ is unitary and antisymmetric for every $\mathbf{k}$, and $\operatorname{Pf}[v(\mathbf{k})]=\operatorname{Pf}[v(C_4T\mathbf{k})]^*\det[B(\mathbf{k})]$. We can make a gauge transformation to make $\operatorname{Pf}[v(\mathbf{k})]=1$ (and thus $\det[B(\mathbf{k})]=1$) for all
 $\mathbf{k}$, and $M(\mathbf{K})$ and $v(\mathbf{K})$ are linked together at $\mathbf{K}\in K^4$ through
\bea
M_{mn}(\mathbf{K})&=&\bra{u_m(\mathbf{K})}\Theta \ket{u_n(\mathbf{K})}
=\bra{u_m(\mathbf{K})} \tilde S^2 \Theta\ket{u_n(\mathbf{K})} \nonumber\\
&=&\eta_{m}(\mathbf{K})\bra{u_m(\mathbf{K})}\tilde S \Theta\ket{u_n(\mathbf{K})} \nonumber\\
&=&\eta_{m}(\mathbf{K})v_{mn}(\mathbf{K})
\label{Mvrelate}
\eea
Hence $\det[M(\mathbf{K})]=\prod_{i=1}^{2N}\eta_i(\mathbf{K})\det[v(\mathbf{K})]$.  Because $\tilde S$ commutes with $C_4T$, bands in the same Kramers pair share identical $\tilde S$ eigenvalues, thus $\prod_{i=1}^{2N}\eta_i(\mathbf{K})=\prod_{r=1}^N\ (\eta_r^\mathrm{I}(\mathbf{K}))^2$ where $\eta_r^\mathrm{I}$ is $\tilde S$ eigenvalue for each Kramers pair. Therefore
\be
\det[M(\mathbf{K})]=\operatorname{Pf}[M(\mathbf{K})]^2=\left( \prod_{r=1}^N\ \eta_r^\mathrm{I}(\mathbf{K}) \operatorname{Pf}[v(\mathbf{K})]   \right)^2
\label{Mvdet}
\ee
Eq.(\ref{Mvrelate}) shows that $\operatorname{Pf}[M(\mathbf{K})]$ is a polynomial of $v_{mn}$ and $\eta_r^\mathrm{I}$, and the square of this polynomial must equal to Eq.(\ref{Mvdet}). When all $\eta_r^\mathrm{I}=1$ we should have $M=v$ and $\operatorname{Pf}[M]=1$. These conditions uniquely fix the polynomial to be $\operatorname{Pf}[M(\mathbf{K})]=\prod_{r=1}^N\ \eta_r^\mathrm{I}(\mathbf{K}) \operatorname{Pf}[v(\mathbf{K})]=\prod_{r=1}^N\ \eta_r^\mathrm{I}(\mathbf{K})$. Since $\det[B]=1$ in this gauge, from Eq.(7) in the main text we have
\be
(-1)^{2P_3}=\prod_{\mathbf{K}\in K^4}\prod_{r=1}^N \eta_r^{\mathrm{I}}(\mathbf{K})
\ee

\section{Tight binding model with $\Theta\ne T$  }

Here we consider another model to show that $\Theta=(C_4T+C_4^{-1}T)/\sqrt{2}$ is the correct operator to use when TRS is broken but $C_4T$ symmetry remains. In this model $C_4=\tau_z e^{i\frac{\pi}{4}\sigma_z}$, $T=-i\tau_0\sigma_yK$ so that $\Theta=\tau_z\sigma_0 T$. Comparing with the model given in the main text where the matrix form for $\Theta$ and $T$ are identical, in this model $\Theta$ and $T$ are manifestly distinct. The Hamiltonian for this model is
\bea
\tilde{H}(\mathbf{k})&=&(\cos k_x+\cos k_y+\cos k_z-2)\tau_z\sigma_0  \nonumber\\
&&+q_1 \sum_{i=x,y} \sin k_i\tau_x\sigma_i+m \tau_x\sigma_z
\label{Hgapped}
\eea
The mass term $m$ opens a full gap in the system. It preserves $C_4T$ but breaks $T$ and $C_4$ separately. Since this mass term is constant, $P_3$ should be trivial \cite{Calugaru2019}, and our open boundary calculation shows there is no hinge states as in Fig.(\ref{gapped}) hence should be trivial. We apply Eq.(13) and (15) in the main text and indeed get $P_3=0$, as predicted. However if we replace $\Theta$ by $T$ in the implementation of Eq.(13) and (15), we find that it will give the incorrect result. Therefore we should stick to $\Theta$ rather than $T$ when time reversal symmetry is broken.

\begin{figure}
\includegraphics[width=3.0in]{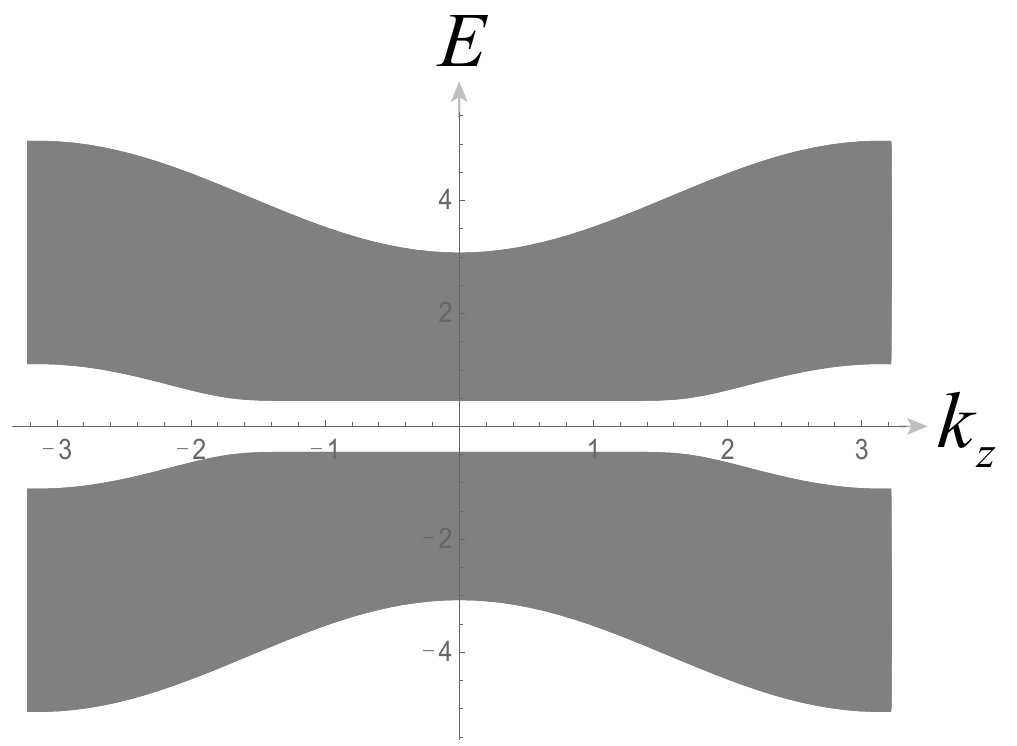}
\caption{ The spectra of $\tilde{H}$ in Eq.(\ref{Hgapped}) as a function of $k_z$ with open boundary condition along $k_x$ and $k_y$ with parameter $q_1=1,m=0.5$. There is no gapless hinge states for $\tilde{H}$, suggesting $P_3=0$.   }
\label{gapped}
\end{figure}

\section{ General construction of the $\Theta$ operator beyond $C_4T$-invariant systems }

In this section we show that the Pfaffian formalism can also be generalized beyond $C_4T$-invariant systems, as long as the system has Kramers-like degeneracy, which means at high symmetry momenta $K$ in the Brillouin zone there is a two-fold degeneracy enforced by an anti-unitary symmetry $\Theta$ so that $\ket{u_n(\mathbf{K})}$ is degenerate with $\Theta\ket{u_n(\mathbf{K})}$. Then the Pfaffian invariant can be defined through the anti-symmetric matrix $M_{mn}(\mathbf{k})=\bra{u_m(\mathbf{k})} \Theta\ket{u_n(\mathbf{k})}$. This requires $\Theta$ to be anti-unitary and $\Theta^2=-1$. Among all the magnetic group elements obtained by a point group operator times $T$, this Kramers-like degeneracy is protected by $C_nT$ with $n=1,3,4$ and $IT$ where $C_n$ is n-fold rotation and $I$ is space inversion. For $C_2T$, $C_6T$ and $\sigma T$ where $\sigma$ is mirror reflection, the system does not have Kramers-like degeneracy any more, because $(C_2T)^2=(\sigma T)^2=+1$ and $(C_6T)^3=-C_2T$. For systems with Kramers-like degeneracy, in $C_4T$-symmetric case $\Theta=\frac{1}{\sqrt{2}}\left( C_4T-(C_4T)^{-1} \right)=\frac{1}{\sqrt{2}}\left( C_4T+C_4^{-1}T \right)$ as given in the main text. For the other cases, the $\Theta$ operator can be constructed similarly. For $n=1$, $\Theta=\frac{1}{2}(T-T^{-1})$; for $n=3$, since $C_3^3=-1$, we can define $\Theta$ as the linear combination $\Theta=\frac{2}{3}\left( C_3T -(C_3T)^{-1}+\frac{1}{2}(C_3T)^3 \right)$ so that $\Theta^2=-1$; for systems with $IT$ symmetry, $\Theta=\frac{1}{2}\left( IT-(IT)^{-1} \right)$. This definition of $\Theta$ provides a generalization of the Pfaffian formalism beyond $C_4T$-invariant systems.

\section{Relationship to quadrupole moment and dipole pumping process  }

In this section we relate our $C_4T$-symmetric insulator to the dipole pumping process discussed in Ref.~\onlinecite{Benalcazar2017,Benalcazar2017b}, which is summarized as follows. Suppose a 2D insulator $H_t(k_x,k_y)$ evolves by a cycle as the adiabatic parameter $t$ changes by $2\pi$. The Hamiltonian at $t=2\pi$ is identical to that at $t=0$. For the quadrupole moment to be well-defined, during the whole process the energy bands and Wannier bands should remain gapped, and the 2D insulator should have a symmetry that guarantees a vanishing bulk polarization, such as $C_2$. At $t=0$ or $\pi$, the 2D insulator has a symmetry such as $C_4$ that quantizes quadrupole moment $q_{xy}$ to $0$ or $1/2$, and at general $t$ the $C_4$ symmetry is broken, allowing the quadrupole to take non-quantized values. Since $q_{xy}$ is defined module 1, if the change of $q_{xy}$ has a winding during the period in which $t$ changes by $2\pi$, the net effect of this cycle is to pump a quantized dipole in a direction perpendicular to it from one edge of the insulator to the other edge. During this dipole pumping process the Wannier-sector polarization $p_y^{\nu_x}$ (defined in Eq.(6.16)~\cite{Benalcazar2017b}) has a winding so that it changes by 1. Here $\nu_x=\nu_x^+$ or $\nu_x^-$ represents different Wannier sectors. If we replace $t$ by $k_z$, the resulting 3D insulator $H(k_x,k_y,k_z)$ has chiral hinge states, and the nontrivial dipole pumping process can be captured by the Wannier-band Chern number $n_{yz}^{\nu_x}$ (defined in Eq.(6.58)~\cite{Benalcazar2017b}).

Now we show that the $C_4T$-symmetric insulator discussed in the main text can be understood by this dipole pumping process, with the only change that the symmetry that quantizes $q_{xy}$ at $k_z=0$ and $\pi$ is $C_4T$ rather than $C_4$. Since $T$ does not change the charge distribution in real space, $C_4T$ can quantize $q_{xy}$ in the same way as $C_4$ when $k_z=0$ or $\pi$. Since $C_2=-(C_4T)^2$, the $C_2$ symmetry in this system also enforces a vanishing bulk polarization for quadrupole to be well-defined at every $k_z$. Since the side surfaces are gapped, the Wannier bands are gapped as well at every $k_z$. Therefore the conditions for dipole pumping are met in this system. We calculate $p_y^{\nu_x}$ as a function of $k_z$ in our model given by Eq.(16) in the main text. We find that when $P_3$ calculated by the Pfaffian formula is nontrivial, $p_y^{\nu_x}$ has a winding with $k_z$, as shown in Fig.(\ref{pwinding}). This winding already shows the Wannier-band Chern number is nontrivial. Our direct calculation of Wannier-band Chern number does give $n_{yz}^{\nu_x^+}=1$. We also calculated the corner charge $Q^{cor}$ for the 2D insulator at each fixed $k_z$ under open boundary condition along both $x$ and $y$. We confirmed that the sign of $Q^{cor}$ at each corner has a quadrupolar fashion, and its value changes from $1/2$ to $-1/2$ during a pumping cycle, indicating a quantized dipole has been pumped during this adiabatic process. Therefore the higher-order topological insulator characterized by $P_3$ is consistent with the dipole pumping picture.

The relationship between the magneto-electric polarization $P_3$ and higher-order topology can also be understood from adiabatic connectivity. For a system with $C_4 T$ symmetry, if we adiabatically deform the Hamiltonian such that the time-reversal (and $C_4$) symmetry is recovered, the system with nontrivial $P_3$ will evolve to a first-order TI with gapless surface states. In the presence of a mass term that breaks both $C_4$ and $T$ but preserves $C_4 T$, the surfaces are gapped, but the mass term changes sign and vanishes at the hinges, resulting in gapless hinge modes that characterize the higher-order topology.

\begin{figure}
\includegraphics[width=3.0in]{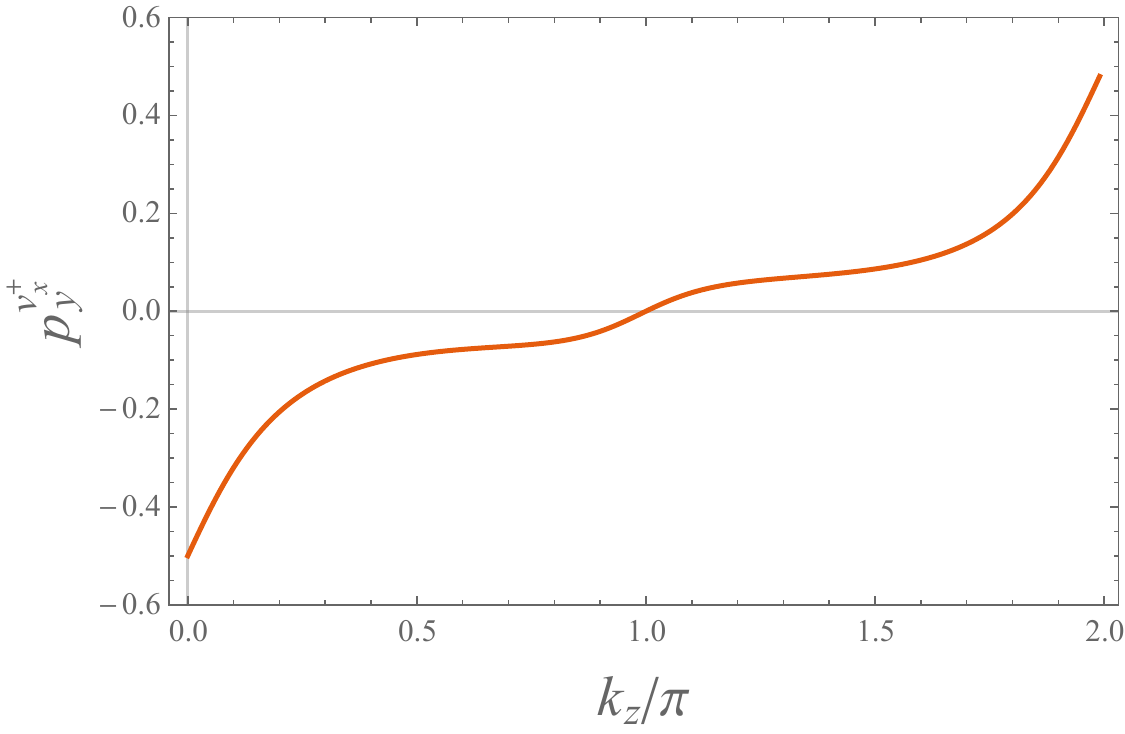}
\caption{ Wannier-sector polarization $p_y^{\nu_x^+}$ as a function of $k_z$ for Hamiltonian Eq.(16) in the main text with parameter $p=0.5,\ q_1=1,\ q_2=0.2,\ q_3=0.3$. $p_y^{\nu_x}$ is defined module 1 within $(-1/2,1/2]$ for $\nu_x=\nu_x^+$ or $\nu_x^-$, and the winding of $p_y^{\nu_x}$ indicates the existence of chiral hinge states.   }
\label{pwinding}
\end{figure}

\end{widetext}

\end{document}